\documentclass[12pt]{article}

\pdfoutput=1

\usepackage{colonequals}
\usepackage{color}
\usepackage{dsfont}
\usepackage{epsfig, palatino}
\usepackage{pstricks,pst-node,pst-tree}
\usepackage{epic}
\usepackage{mathrsfs}
\usepackage{ae} 
\usepackage[T1]{fontenc}
\usepackage[ansinew]{inputenc}
\usepackage{amsmath}
\usepackage{amssymb}
\usepackage{graphicx}
\usepackage[colorlinks=true,linkcolor=blue,citecolor=blue,urlcolor=blue]{hyperref}
\usepackage{cite}
\usepackage{hyperref}
\usepackage{wasysym}
\usepackage{varioref}
\usepackage{makeidx}
\usepackage[english]{babel}
\usepackage{simplewick}
\usepackage{array}
\usepackage{multirow}

\usepackage[font={small}]{caption}

\renewcommand\]{\end{equation}}
\renewcommand\[{\begin{equation}}

\makeindex

\newcommand\blfootnote[1]{%
  \begingroup
  \renewcommand\thefootnote{}\footnote{\hspace{-6mm}#1}%
  \addtocounter{footnote}{-1}%
  \endgroup
}

\begin{document}

\thispagestyle{empty}

\renewcommand{\thefootnote}{\fnsymbol{footnote}}
\setcounter{page}{1}
\setcounter{footnote}{0}
\setcounter{figure}{0}


\title{Large scalar gaps in 2D CFTs with generalized polynomials}

\date{}

\maketitle

\begin{center}

\textrm{Renato G. F. Souza$^\text{\tiny 1,\tiny 2,\tiny 3}$}

\vspace{1.2cm}
\footnotesize{\textit{
$^\text{\tiny 1}$ICTP South American Institute for Fundamental Research, IFT-UNESP, S\~ao Paulo, SP Brazil 01440-070 \\
$^\text{\tiny 2}$Perimeter Institute, 31 Caroline Street North, Waterloo, ON, N2L 2Y5, Canada   \\
$^\text{\tiny 3}$Department of Physics and Astronomy, University of Waterloo,
Waterloo, Ontario, Canada, N2L 3G1 }}

\blfootnote{\tt  rgnato61@gmail.com}
\\ \vspace{1.2cm}
\footnotesize{\textit{
}
\vspace{4mm}
}

\end{center} 

\begin{abstract}
    We present an analytic way of writing simple crossing symmetric expressions and use them to search for unitary 4-point functions in 2D CFTs. We've applied our method for a class of functions we called generalized polynomials to achieve large gaps for operators with integer scaling dimension less or equal to 18.
\end{abstract}

\newpage

\tableofcontents

\newpage

\section{Introduction} \label{sec:1}

The CFT bootstrap program started in the 70's with the works \cite{FERRARA1973161,Polyakov:1974gs}. Using the Virasoro algebra they solved every unitary 2D CFT with central charge $c<1$. However they were unable to extend their methods to higher dimensions or larger central charge. 

The recent revival of the bootstrap program caused by \cite{Rattazzi_2008} was due to its novel ways of analyzing crossing symmetry. Not only it was able to be replicated in arbitrary dimensions but was also extended to similar feats of the original in special cases, as the $3D$ Ising and the $O\left(N\right)$ models, \cite{El_Showk_2012,Kos_2014,Kos_2015}. With their new formulation we are able to show for which values of the CFT data there is no unitary CFT.

Our method is dual to theirs, we demonstrated how to construct simple unitary 4-point functions using the generalized polynomials. These will be points that cannot be excluded by their methods. An important distinction between the two is that this is purely analytical. We won't be able to find every unitary CFT. However, as will be shown in section \ref{sec:4}, the generalized polynomials cover a substantial region of the space for small scaling dimension.

This work is divided as follows, in section \ref{sec:2} we review the numerical bootstrap approach to motivate why starting with crossing symmetric expressions is interesting. In section \ref{sec:4} we introduce and analyze the generalized polynomials. Lastly, in \ref{sec:5} we present a computational method to find large scalar gaps.

The appendices cover the details not present in the main text, in \ref{ap:A} we discuss a special case of generalized polynomials and the relation between crossing symmetry and a more general basis. Appendix \ref{ap:B} presents the properties of the hypergeometric functions in the OPE expansion and proves the important statements from the main text.

\section{Reviewing the bootstrap equation and crossing symmetry} \label{sec:2}

Before starting it would be good to review some points. For a more in depth review check \cite{Poland:2018epd,Rychkov:2016iqz,simmonsduffin2016tasi}. The two main elements in the CFT bootstrap program are the OPE and crossing symmetry. We can use the OPE to write all correlation functions in terms of the CFT data of primary operators, $\left\{\Delta_{i},\lambda_{ijk}\right\}$, and we can use crossing symmetry to relate different orderings in a correlator. A 4-point function of identical scalars $\phi$ is give by

\begin{align}
    &\mathcal{G}\left(u,v\right)=\underset{J,\Delta}{\sum} \left(\lambda_{\phi\phi\mathcal{O}}\right)^{2} G^{\left(D\right)}_{J_{\mathcal{O}},\Delta_{\mathcal{O}}}\left(u,v\right), \\
    \label{eq:02} & \mathcal{G}\left(u,v\right)=\mathcal{G}\left(\frac{u}{v},\frac{1}{v}\right)=\left(\frac{u}{v}\right)^{\Delta_\phi}\mathcal{G}\left(v,u\right).
\end{align}

In the numerical bootstrap the conformal block decomposition is applied to the crossing relations to obtain the bootstrap equation. This new manifestly unitary constraint is analyzed numerically with functionals to bound the CFT data

\begin{align} \label{eq:06}
    \underset{J,\Delta}{\sum}\left(\lambda_{\phi \phi \mathcal{O}}\right)^{2}\left( \frac{G^{\left(D\right)}_{J_\mathcal{O},\Delta_{\mathcal{O}}}\left(u,v\right)}{u^{\Delta_{\phi}}}-\frac{G^{\left(D\right)}_{J_\mathcal{O},\Delta_{\mathcal{O}}}\left(v,u\right)}{v^{\Delta_{\phi}}}\right)=\underset{J,\Delta}{\sum}\left(\lambda_{\phi \phi \mathcal{O}}\right)^{2} F_{J,\Delta}^{\Delta_{\phi}}\left(u,v\right)=0, ~~~~ \left(\lambda_{\phi \phi \mathcal{O}}\right)^{2}\geq0.
\end{align}

In practice we cannot expect to find an exact unitary 4-point function from this equation, but we can from the crossing relations $\left(\ref{eq:02}\right)$. If we construct a crossing symmetric expression and then show that it satisfy unitarity we would have a point in the data space that cannot be excluded without new constrains. In the optimal case we would be able to pinch the boundary between the excluded and allowed region, similar to \cite{2020}, however this is not what we have now. Although not optimal this still allow us to find interesting results.

Assuming we already have a crossing symmetric expression, whether a closed form or an approximation, the challenge becomes finding when they are unitary. The conformal blocks are closely related to the analytic structure of 4-point functions at the point $u=0$, however crossing symmetry is related to series at the point $u=v=1$. This loss of information about $u=0$ is where most of the difficulty resides. In appendix \ref{ap:A} we show this in a simple example.

To bypass this loss we search for solutions of crossing symmetry that behaves similar to a conformal block close to $u=0$. The conformal blocks behave as $G^{\left(D\right)}_{J,\Delta} \sim \mathcal{N}_{J,\Delta} u^{\frac{\Delta-J}{2}}$, $\left(u\xrightarrow[]{}0 ~~ v\xrightarrow[]{}1\right)$, with our normalization shown in \ref{ap:B}. We use a basis of functions $\left\{f_{i}\left(u,v\right)\right\}$ that for small $u$ it behaves as a sum of power laws, $f_{i}\left(u,v\right) \sim \sum_{j} \mathcal{C}_{i, j}\left(v\right) u^{p_{i, j}}$ and are flexible enough to work for any scaling dimension in $\left(\ref{eq:02}\right)$. We will start with the simplest basis, the generalized polynomials, and show that it already has interesting results.

\section{Generalized polynomials} \label{sec:4}

The generalized polynomials are polynomial-like functions with arbitrary powers of $u$ and $v$ and with a finite number of terms to keep crossing symmetry under control. They are given by the ansatz

\begin{align} \label{eq:12}
    & \mathcal{G}\left(u,v\right) = \frac{1}{v^{\Delta_{\phi}}}\underset{i=1}{\overset{N<\infty}{\sum}}~ \xi_{i} ~ u^{a_{i}} v^{b_{i}} = \underset{i=1}{\overset{M}{\sum}}~ E_{i} \left(a_{i},b_{i},c_{i}\right), \\
    \label{eq:13} \left(a,b,c\right) &= \frac{u^{a} v^{b}+u^{a}v^{c}+u^{b}v^{a}+u^{b}v^{c}+u^{c}v^{a}+u^{c}v^{b}}{N_{a,b,c} ~ v^{\Delta_{\phi}}}, ~~ a+b+c=2\Delta_{\phi}, \\
     N_{a,b,c} &= \left\{\begin{tabular}{c}
         1 if $a\neq b$, $b\neq c$, $c\neq a$, \\
         2 if $a=b\neq c$, $b=c\neq a$, $c=a\neq b$, \\
         6 if $a=b=c$.
    \end{tabular}\right.
\end{align}

From the polynomial-like basis $\left\{u^{a_{i}} v^{b_{i}-\Delta_{\phi}}\right\}$ we created a new crossing symmetric basis $\left\{\left(a_{i},b_{i},c_{i}\right)\right\}$ which we can use to construct our correlators. Using the power law behavior at $u=0$ we can predict what is the form of their conformal block decomposition. By matching the powers in the conformal blocks with those in the polynomial basis we find that the correct series should have the following structure

\begin{align} \label{eq:15}
    u^{p} {v^{q}} \sim \underset{J,n}{\sum} A^{\left(D\right)}_{J,n} ~ G_{J,2p+J+n}^{\left(D\right)}\left(u,v\right).
\end{align}

We can see that the series generated by two power laws $u^{p}$ and $u^{q}$ won't overlap unless their powers differ by an integer or half-integer.

Unlike unitarity, crossing symmetry do not depend on the number of dimensions, if $D\geq2$, as such expressions $\left(\ref{eq:12}\right)$ and $\left(\ref{eq:15}\right)$ are general results. In
 \ref{ap:B} we give all the properties needed to show that the correct conformal block decomposition for general $D$ has the form

\begin{align} \label{eq:16}
    \frac{u^{p}\left(v^{q}+v^{r}\right)}{v^{\Delta_{\phi}}} = \underset{m\geq n}{\sum} \frac{1+\left(-1\right)^{n+m}}{n! m!} F^{n,m}_{p,D}\left(\left|\frac{q-r}{2}\right|\right) G^{\left(D\right)}_{m-n,2p+n+m}\left(u,v\right),
\end{align}

where $F_{p,D}^{n,m}\left(t\right)$ is a polynomial in $t$. For the rest of the main text we set $D=2$.

The relation between the $2D$ and $1D$ conformal blocks, $G_{J,\Delta}\sim k_{\frac{\Delta+J}{2}}k_{\frac{\Delta-J}{2}}$, allow us to use the results found in \cite{Hogervorst_2017} to calculate $F_{p,2}^{n,m}\left(t\right)$ exactly. Both the result and the relation between conformal blocks are written in the Dolan-Osborn variables, so we rewrite the functions $u^{p}v^{q}$ as $ \left(z^{p}\left(1-z\right)^{q}\right)\left(\bar{z}^{p}\left(1-\bar{z}\right)^{q}\right)$ and use \cite{Hogervorst_2017} to obtain

\begin{align} \label{eq:18}
    z^{p}\left(1-z\right)^{q} = \underset{n=0}{\overset{\infty}{\sum}} \dfrac{\left(-1\right)^{n} \left(p\right)^{2}_{n}}{n! \left(2p-1+n\right)_{n}} {}_3F_{2} \left(\left.\begin{array}{c}
    \begin{array}{ccc}
        -n, & 2p-1+n, & -q \\
    \end{array}{}\\ \begin{array}{cc}
        p, & p \\
    \end{array}{} \end{array}{} \right| 1\right) k_{p+n}\left(z\right).
\end{align}

These hypergeometric function have multiple properties we can exploit, to make them simpler to express we'll rewrite them as

\begin{align} \label{eq:19}
    f_{p}^{n}\left(t\right) &= \dfrac{\left(p\right)^{2}_{n}}{\left(2p-1+n\right)_{n}} {}_3F_{2} \left(\left.\begin{array}{c}
    \begin{array}{ccc}
        -n, & 2p-1+n, & \frac{p}{2}-t \\
    \end{array}{}\\ \begin{array}{cc}
        p, & p \\
    \end{array}{} \end{array}{} \right| 1\right),
\end{align}

which yields

\begin{align}
    z^{p}\left(1-z\right)^{q} &= \underset{n=0}{\overset{\infty}{\sum}} \dfrac{\left(-1\right)^{n} }{n!} f_{p}^{n}\left(\frac{p}{2}+q\right) k_{p+n}\left(z\right),\\ 
    \label{eq:20} F_{p,2}^{n,m}\left(t\right) &= 2^{m-n}f_{p}^{n}\left(t\right)f_{p}^{m}\left(t\right).
\end{align}

The first main property of $f_{p}^{n}\left(t\right)$ is its parity, $f_{p}^{n}\left(-t\right)=\left(-1\right)^{n}f_{p}^{n}\left(t\right)$, so we only need to analyze the region $t\geq0$. Second $f$ is a continuous Hahn polynomial, meaning they have a simple recurrence relation

\begin{align} \label{eq:22}
    f_{p}^{n+1}\left(t\right)=t f_{p}^{n}\left(t\right)+\dfrac{n\left(n+p-1\right)^{2} \left(n+2p-2\right)}{4\left(2n+2p-1\right)\left(2n+2p-3\right)} f_{p}^{n-1}\left(t\right).
\end{align}

Using induction together with the initial conditions $f_{p}^{0}\left(t\right)=1$ and $f_{p}^{1}\left(t\right)=t$ and $p\geq0$ we prove that $f_{p}^{n}\left(t\right)\geq0$. The only step left is to prove $\left(\ref{eq:16}\right)$ for $D=2$ and $a+b+c=2\Delta_{\phi}$

\begin{align} \label{eq:23}
    & \nonumber \frac{u^{p}\left(v^{q}+v^{r}\right)}{v^{\frac{p+q+r}{2}}} = \underset{m\geq n}{\sum} \frac{\left(-1\right)^{n+m}}{n! m!} \left( F_{p,2}^{n,m}\left(\frac{q-r}{2}\right)+F_{p,2}^{n,m}\left(\frac{r-q}{2}\right) \right) G^{\left(2\right)}_{m-n,2p+n+m}\left(u,v\right) \\
    & \nonumber = \underset{m\geq n}{\sum} \frac{\left(-1\right)^{n+m}}{n! m!} \left( F_{p,2}^{n,m}\left(\left|\frac{q-r}{2}\right|\right)+F_{p,2}^{n,m}\left(-\left|\frac{q-r}{2}\right|\right) \right) G^{\left(2\right)}_{m-n,2p+n+m}\left(u,v\right) \\
    & = \underset{m\geq n}{\sum} \frac{\left(-1\right)^{n+m}}{n! m!}\left(1+\left(-1\right)^{n+m}\right)  F_{p,2}^{n,m}\left(\left|\frac{q-r}{2}\right|\right)  G^{\left(2\right)}_{m-n,2p+n+m}\left(u,v\right).
\end{align}

Although all these OPE coefficients are positive they are not sufficient to guarantee unitarity.

\subsection{Unitary generalized polynomials and the scalar gap}

For a 4-point function of identical real scalars we expect the identity to appear in the OPE, but according to $\left(\ref{eq:23}\right)$ this is not the case when all parameters in $\left(a,b,c\right)$ are non-zero. Since we've only assumed identical scalars we don't know whether $\phi$ is null or it is charged, in any case, we cannot call $\mathcal{G}$ unitary. Still it is possible for any $\left(a,b,c\right)$ to generate an unitary 4-point function. Because of the positivity of $\left(\ref{eq:23}\right)$ the sum $\mathcal{G}_{\mathbf{1}}+\left(a,b,c\right)$ will be unitary if $\mathcal{G}_{\mathbf{1}}$ is unitary. For us $\mathcal{G}_{\mathbf{1}}$ will be just another generalized polynomial. Bellow we plot the scalar gap of every possible unitary 4-point function we can guarantee exists from $\left(\ref{eq:23}\right)$.

\begin{figure}[h]
    \centering
    \includegraphics[scale=0.5]{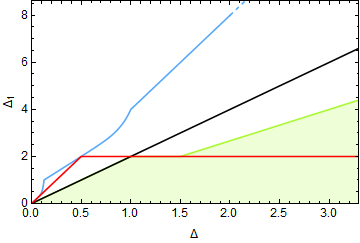}
    \caption{The blue curve shows an approximation to the bound given by the numerical bootstrap, the other curves and regions are the places we can find unitary solutions using only $\left(\ref{eq:23}\right)$.}
    \label{fig:01}
\end{figure}

There are two cases that stand out in the plot, these correspond to the generalized free boson, $\left(0,\Delta_{\phi},\Delta_{\phi}\right)$, as the black line and the cosine operators in the free boson, $\left(0,0,2\Delta_{\phi}\right)$, as the red curve.

Until now all we've analyzed is equivalent to a basis with a single element $\left\{\left(a,b,c\right)\right\}$. For a general basis we cannot find a complete set of solutions like we did previously, however we can still focus in searching for interesting 4-point functions. In our case functions with large scalar gap.

Non-trivial solutions, those containing negative basis coefficients, form the majority of the interesting 4-point functions, yet most basis cannot create them. This is the result of two properties of $\left(\ref{eq:23}\right)$, first, all OPE coefficients are positive and second, the operators in the OPE have twist $\tau=2p+2n$. The first means that whenever a basis coefficient becomes negative there are up to 3 potential series of only negative OPE coefficients. The second severely limits which kinds of terms could be used to counteract the first.

Choosing a basis is, then, an extremely important first step when trying to analyze any property of a CFT. Once one is chosen we need to know which tools we have to say with certain confidence whether an expression is unitary or not. If the number of elements is small we can still use analytical methods to prove unitarity, but for larger basis we will usually rely in the asymptotic expansion of the $f_{p}^{n}\left(t\right)$ polynomials. To test this method we start by analyzing small scaling dimensions. Since we didn't need a large basis to get close to the boundary we used analytical methods to prove their existence.

\begin{figure}[h]
    \centering
    \includegraphics[scale=0.5]{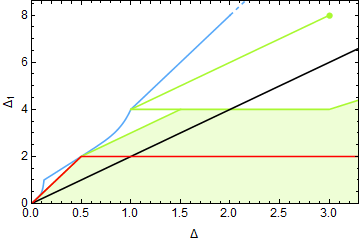}
    \caption{Regions we could prove the existence of unitary solution analytically.}
    \label{fig:02}
\end{figure}

\section{Searching for large gaps} \label{sec:5}

Despite having exact expression for the OPE coefficients there still is a small, but important, hassle in $\left(\ref{eq:19}\right)$, the function $f_{0}^{n}\left(t\right)$ is not well defined. For $n\geq2$ the following redefinition is enough

\begin{align}
    f_{p}^{n}\left(t\right) &= \dfrac{\left(\Gamma\left(p+n\right)\right)^{2}}{\left(2p-1+n\right)_{n}} {}_3 \overset{\sim}{F}_{2} \left(\left.\begin{array}{c}
    \begin{array}{ccc}
        -n, & 2p-1+n, & \frac{p}{2}-t \\
    \end{array}{}\\ \begin{array}{cc}
        p, & p \\
    \end{array}{} \end{array}{} \right| 1\right).
\end{align}

By analyzing $\left(1-z\right)^{t}=1-t~z+\mathcal{O}\left(z^{2}\right)$ it becomes clear that $f_{0}^{0}\left(t\right)=1$ and $f_{0}^{1}\left(t\right)=t$. With every value of $f_{p}^{n}\left(t\right)$ defined it will be useful to think of $\frac{f_{p}^{n}}{n!}\equiv0$ for $n<0$. As was said, the operators in the OPE for a generalized polynomials have twist $\tau=2p+2n$, this means that these series overlap only when the parameter differ by an integer. We will represent this overlapping series by $\mathcal{E}_{\left[p\right]}$, $\left[p\right]=p-\lfloor p \rfloor$

\begin{align}
    & \mathcal{G} = \underset{\left[p\right]}{\sum} \underset{m\geq n}{\sum} \left(1+\left(-1\right)^{n+m}\right)\mathcal{E}_{\left[p\right]}^{n,m} G^{\left(2\right)}_{m-n,2\left[p\right]+n+m}, \\ 
    \label{eq:28} & \mathcal{E}_{\left[p\right]}^{n,m} = \underset{i}{\sum} 2^{m-n}E_{i}\frac{ f_{\left[p\right]+\lfloor p_{i}\rfloor}^{n-\lfloor p_{i}\rfloor}\left(t_{i}\right)}{\left(n-\lfloor p_{i}\rfloor\right)!}  \frac{f_{\left[p\right]+\lfloor p_{i}\rfloor}^{m-\lfloor p_{i}\rfloor}\left(t_{i}\right) }{\left(m-\lfloor p_{i}\rfloor\right)!}.
\end{align}

No matter which property we try to explore using this method the code should always search for a set of coefficients $E_{i}$ that makes every $\mathcal{E}_{\left[p\right]}^{n,m}$ non-negative. To deal with the infinite number of inequalities we can use the asymptotic form of the $f$ polynomials. By making sure that the asymptotic expression is positive and that $\mathcal{E}_{\left[p\right]}^{n,m}\geq0$ for a sufficient large $n$ and $m$ we would know with confidence whether the expression is unitary.

For our search for large scalar gaps this can be coded by introducing two types of linear constrains, the positivity plus normalization, $\mathcal{E}_{0}^{0,0}=1$ and $\mathcal{E}_{\left[p\right]}^{n,n+J}\geq0$ for $n\leq n_{min}$ and $J\leq J_{min}$, and the gap condition, $\mathcal{E}_{\left[p\right]}^{n,n}=0$ if $0<2\left[p\right]+2n<\Delta_{1}$. Any set of $E_{i}$ that satisfy these conditions could be further tested for $\mathcal{E}_{\left[p\right]}^{n,n+J}\geq0$ when $n\leq n_{max}$ and $J\leq J_{max}$.

For the green points in the plots bellow we've used the basis
 $\mathcal{B}=\left\{\left(a,b,c\right)\left|a,b,c\in \mathbb{Z}_{\geq0} \right\}\right.$. In this special basis the value of $n_{max}$ and $J_{max}$ can be calculated using some analytical results from the continuous Hahn polynomials. This way we've proved that there are no negative OPE coefficients. This is not the biggest possible gap, as is shown in the second plot bellow when adding half-integers to the basis the gap increases.

\begin{figure}[h]

\begin{minipage}[b]{0.48\linewidth}

    \centering
    \includegraphics[scale=0.5]{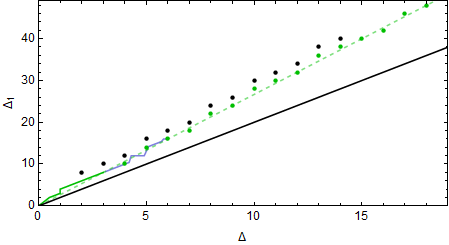}
    \caption{The green dots are new unitary solutions, compared to the last plot, the light blue are possible unitary, we tested until twist and spin 200, the black are points we cannot reach with our current basis, the black line is the generalized free boson and the dashed line is $\frac{8\Delta}{3}$.}
    \label{fig:03}

\end{minipage} 
\hfill
\begin{minipage}[b]{0.48\linewidth}

    \centering
    \includegraphics[scale=0.5]{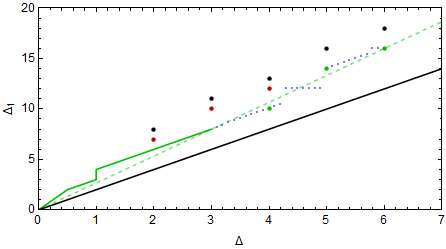}
    \caption{By introducing half-integers parameters to the basis we were able to reach the red points, the black ones are points we cannot reach with this new basis.}
    \label{fig:04}

\end{minipage} 

\end{figure}

In AdS/CFT the large radius limit is a way to recover the flat space-time physics, \cite{Paulos_2017}. In this limit theories like the generalized free boson are interpreted as free particles with their masses proportional to the scaling dimension $\Delta_{\phi}$. The scalar gap of $2\Delta_{\phi}$ can be interpreted the 2-particle bound state have twice the particle mass. Figure \ref{fig:03} shows that for large $\Delta_{\phi}$ there are theories with scalar gap close to $\frac{8\Delta_{\phi}}{3}$. Any theories with the same type of interpretation in this region would have $\frac{8m}{3}$ as the bound state mass of particles of mass $m$. This behavior is not restricted to integers scaling dimension, with a less robust program it was possible to find possible solutions with non-integer scaling dimension connecting these points.

\section{Conclusion}

In this work we presented a method to construct simple unitary 4-point functions. Unlike the dual formalism from \cite{Rattazzi_2008} this is completely analytical, as such we are not able to create every possible 4-point functions. We've analyzed the generalized polynomials trying to maximize the scalar gap.

We used the properties of the hypergeometric polynomials $f_{p}^{n}\left(t\right)$ to study their OPE decomposition and to create a computational method to search for unitary solutions. The main results are given by figure \ref{fig:02} and \ref{fig:03}. Figure \ref{fig:02} shows that despite its simple form the generalized polynomials seems to cover a significant portion of the allowed physical space. Figure \ref{fig:03} shows that even for relative large scaling dimensions it still possible to find unitary 4-point functions with significantly large scalar gap than the generalized free boson. In the large radius limit of AdS/CFT this can be related to the mass of the 2-particle bound state, that is, their bound state would have a mass significantly larger than twice the particle mass.

The results from the program, which can be found at \href{https://github.com/rgnato/Generalized_polynomials}{GitHub}, are valid only for $D=2$, however in \ref{ap:B} we used the results of \cite{Hogervorst_2016,Kaviraj_19} to find recursion relations for the $F_{p,D}^{n,m}\left(t\right)$ polynomials in $D$. Most of the code could be re-purposed to analyze higher dimensional CFTs or other observables, but we may need another way to prove, or at least be confident about unitarity.

\section*{Acknowledgements}

I would like to thank my advisor Pedro Vieira for helping me during this project, for the discussions and for the ideas exchanged. I like to thank both Pedro, Alexandre Homrich and Matheus Fabri for helping check this work. I would also like to thanks the CNPq for the partial financial support with the grant No. 132286/2018-1 during my stay in the IFT-UNESP/ICTP-SAIFR. And during my stay at PI. The work was supported in part by the Natural Sciences and Engineering Research Council of Canada. Research at Perimeter Institute is supported in part by the Government of Canada through the Department of Innovation, Science and Economic Development Canada and by the Province of Ontario through the Ministry of Economic Development, Job Creation and Trade.

\appendix

\section{Notes}\label{ap:A}

In this appendix we'll go through some small details we ignored in the main text. We leave the analysis of the functions $F_{a,D}^{n,m}\left(t\right)$ and $f_{a}^{n}\left(t\right)$ to \ref{ap:B}.

\subsubsection*{The relation between series and functional equations}

In \ref{sec:2} we've mentioned how crossing symmetry can be thought as a series around the symmetric point. To give an example suppose we have the functional equation $f\left(z\right)=-f\left(1-z\right)$, and $f$ is regular at $z=\frac{1}{2}$. The general solution can be written as $f\left(z\right)=\sum_{i=0}^{\infty}c_{i}\left(z-\frac{1}{2}\right)^{2i+1}$.

Even the most general case, $f\left(z\right)=g\left(z\right)f\left(\frac{z+b}{c z-1}\right)$ and $f$ and $g$ are regular at the symmetric point $z_{0}$, will give us the solution $f\left(z\right)=\frac{1}{\sqrt{g\left(z_{0}\right)g\left(\frac{z+b}{c z-1}\right)}}\sum_{i=0}^{\infty}c_{i}\left(\frac{b+z\left(1\pm\sqrt{1+b c}\right)}{b+z\left(1\mp\sqrt{1+b c}\right)}\right)^{2i+\epsilon}$, $\epsilon=\theta\left(-g\left(z_{0}\right)\right)$. The sign $\pm$ is determined by making sure that $z_{0}$ is mapped to $0$.

We can use this to solve the 1D crossing symmetry $f\left(z\right)=\left(\frac{z}{1-z}\right)^{2\Delta}f\left(1-z\right)$

\begin{align}
    f\left(z\right)=\left(\frac{z}{1-z}\right)^{\Delta}\sum_{i=0}^{\infty} c_{i}\left(z-\frac{1}{2}\right)^{2i}.
\end{align}

Unlike the generalized polynomials this expression do approximate every possible 4-point function. Although we lose control over the spectrum it is not a downside. In the generalized polynomials we always input our spectrum, but we don't known what they are for theories on the boundary. For convergence we define $x\left(z\right)=\frac{\sqrt{z}-\sqrt{1-z}}{\sqrt{z}+\sqrt{1-z}}$, mapping the plane into a disk

\begin{align}
    f\left(z\right)=\left(\frac{\left(x\left(z\right)+1\right)^{2}}{\left(x\left(z\right)-1\right)^{2}}\right)^{\Delta}\sum_{i=0}^{\infty} b_{i}\left(x\left(z\right)\right)^{2i}.
\end{align}

Notice, however, that $z=0$ is at the boundary at the convergence region, so it is not clear how to recover the CFT data or prove unitarity from this expression.

\subsubsection*{Expansion with $F_{p,2}^{n,m}\left(0\right)$}

To find the expression in equation $\left(\ref{eq:23}\right)$ it was assumed that $b\neq c$, so that 

\begin{align} 
    \nonumber \frac{u^{a}\left(v^{b}+v^{c}\right)}{v^{\Delta_{\phi}}} = \underset{m\geq n}{\sum} \frac{1+\left(-1\right)^{n+m}}{n! m!}  F_{a,2}^{n,m}\left(\left|\frac{b-c}{2}\right|\right)  G^{\left(2\right)}_{m-n,2a+n+m},
\end{align}

however, by the normalization chosen for the $\left(a,b,c\right)$, if $b=c$ it becomes

\begin{align}
    & \frac{u^{a} v^{b}}{v^{\Delta_{\phi}}} = \underset{m\geq n}{\sum} \frac{\left(-1\right)^{n+m}}{n! m!} F_{a,2}^{n,m}\left(0\right)  G^{\left(2\right)}_{m-n,2a+n+m}\left(u,v\right).
\end{align}

Since $f_{a}^{n}\left(0\right)$ is $0$ whenever $n$ is odd, $F_{a,2}^{n,m}\left(0\right)$ will be $0$ whenever $n$ or $m$ is odd, as such $F_{a,2}^{n,m}\left(0\right)=\frac{1+\left(-1\right)^{n+m}}{2}F_{a,2}^{n,m}\left(0\right)$. As we'll prove the parity of $F_{p,2}^{n,m}$ is true for every $F_{p,D}^{n,m}$, so we can use it to write

\begin{align}
    & \frac{u^{a} v^{b}}{v^{\Delta_{\phi}}} = \underset{m\geq n}{\sum} \frac{1}{2} \frac{1+\left(-1\right)^{n+m}}{n! m!} F_{a,D}^{n,m}\left(0\right)  G^{\left(D\right)}_{m-n,2a+n+m}\left(u,v\right).
\end{align}

\section{Properties of the polynomials $f_{p}^{n}$ and $F_{p,D}^{n,m}$}\label{ap:B}

Since both $f_{p}^{n}\left(t\right)$ and $F_{p,D}^{n,m}\left(t\right)$ play a prominent role in the unitarity of the generalized polynomials it would be natural to analyze them more deeply.

\subsection{Parity, recursion in $n$ and asymptotic behavior of $f_{p}^{n}$}

All properties of the continuous Hahn polynomials used in this work can be found in the book \cite{book}. Its parity can be found from one of the generating functions

\begin{align}
        {}_{1}F_{1}\left(\left.\begin{array}{c}
         \begin{array}{c}
             \frac{p}{2}-t \\
             p
         \end{array}{}
    \end{array}\right|x\right)
    {}_{1}F_{1}\left(\left.\begin{array}{c}
         \begin{array}{c}
             \frac{p}{2}+t \\
             p
         \end{array}{}
    \end{array}\right|-x\right)= \underset{n=0}{\overset{\infty}{\sum}}\frac{\left(n+2p-1\right)_{n}f_{p}^{n}\left(t\right)}{n!\left(p\right)^{2}_{n}} x^{n}.
\end{align}

The generating function is invariant under the map $\left\{t\mapsto-t,x\mapsto-x\right\}$, since the coefficients of $x^{n}$ needs to be identical in both cases it proves $f_{p}^{n}\left(t\right)$, and $F_{p,2}^{n,m}\left(t\right)$, parity.

For simple basis a useful relation is the one given by the forward shift operator

\begin{align}
    f_{p}^{n}\left(t\right)-f_{p}^{n}\left(t-1\right)=n f_{p+1}^{n-1}\left(t-\frac{1}{2}\right).
\end{align}

Another useful relation for small basis comes from the recursion relation $\left(\ref{eq:22}\right)$. Using the following inequality

\begin{align}
    f_{p}^{n+1}\left(t_{1}\right)-f_{p}^{n+1}\left(t_{2}\right) & \geq t_{2}\left(f_{p}^{n}\left(t_{1}\right)-f_{p}^{n}\left(t_{2}\right)\right)+\\
    & \nonumber \dfrac{n\left(n+p-1\right)^{2} \left(n+2p-2\right)}{4\left(2n+2p-1\right)\left(2n+2p-3\right)}\left(f_{p}^{n-1}\left(t_{1}\right)-f_{p}^{n-1}\left(t_{2}\right)\right),
\end{align}

which
 we can use to prove $f_{p}^{n}\left(t_{1}\right)\geq f_{p}^{n}\left(t_{2}\right)$ if $t_{1}\geq t_{2}\geq0$. For very simple basis these two conditions are sufficient to prove unitarity.

The asymptotic relation for $f_{p}^{n}\left(t\right)$ can be obtained from its Rodrigues-type formula

\begin{align}
    f_{p}^{n}\left(t\right)=\frac{\underset{k=0}{\overset{n}{\sum}}\left(-1\right)^{k} \binom{n}{k} \left(\Gamma\left(\frac{p}{2}+n-k+t\right)\right)^{2} \left(\Gamma\left(\frac{p}{2}+k-t\right)\right)^{2}}{\left(n+2p-1\right)_{n}\left(\Gamma\left(\frac{p}{2}+t\right)\right)^{2} \left(\Gamma\left(\frac{p}{2}-t\right)\right)^{2}}.
\end{align}

By rewriting $f_{p}^{n}\left(t\right)$ as the sum $\sum_{k=0}^{\lfloor\frac{n}{2}\rfloor} \phi_{k}\left(n\right)$, with

\begin{align}
    \phi_{k}\left(n\right) = \dfrac{\left(-1\right)^{k}\binom{n}{k}}{\left(n+2p-1\right)_{n}}\left( \dfrac{\left(\Gamma\left(\frac{p}{2}+n-k+t\right) \Gamma\left(\frac{p}{2}+k-t\right)\right)^{2}}{\left(\Gamma\left(\frac{p}{2}+t\right) \Gamma\left(\frac{p}{2}-t\right)\right)^{2}} \right. + \\
    \nonumber \left. \left(-1\right)^{n} \dfrac{\left(\Gamma\left(\frac{p}{2}+n-k-t\right) \Gamma\left(\frac{p}{2}+k+t\right)\right)^{2}}{\left(\Gamma\left(\frac{p}{2}+t\right) \Gamma\left(\frac{p}{2}-t\right)\right)^{2}} \right),
\end{align}

the order $\mathcal{O}\left(\frac{n^{2t+p-\frac{3}{2}-k}}{2^{2n+2p}}\right)$ asymptotic behaviour can be calculated by using all $\phi$ up to $\phi_{k}$. For $t\geq0$ the general form of the asymptotic behavior is

\begin{align} \label{eq:55}
    \dfrac{f_{p}^{n}\left(t\right)}{n!} \sim \dfrac{\sqrt{2 \pi} ~ n^{2t+p-\frac{3}{2}}}{2^{2n+2p-\frac{3}{2}}\left(\Gamma\left(\frac{p}{2}+t\right)\right)^{2}}\left(1+\mathcal{O}\left(n^{-1}\right)+\left(-1\right)^{n}\mathcal{O}\left(n^{-4t}\right)\right)\equiv\frac{Af_{p}^{n}\left(t\right)}{n!}.
\end{align}

\subsection{Recursions in $D$ of $F_{p,D}^{n,m}$}

Unlike the polynomials $f_{p}^{n}\left(t\right)$ which have multiple properties, the polynomials $F_{p,D}^{n,m}\left(t\right)$ seems to only have recursion in $D$ and parity. The recursions in $D$ can be calculated using the dimension reduction for the conformal blocks, for $D$ and $D+2$ we'll use section 4.4.3 of \cite{Kaviraj_19} and for $D$ and $D+1$ we'll use \cite{Hogervorst_2016}.

All relations exist because we can expand the same expression in terms of conformal blocks of different dimensions

\begin{align} \label{eq:34}
    u^{p} v^{t-\frac{p}{2}} & = \underset{m\geq n}{\sum} \frac{\left(-1\right)^{n+m}}{n! m!} F_{p,2}^{n,m}\left(t\right)  G_{m-n,2p+n+m}^{\left(2\right)}
    = \underset{m\geq n}{\sum} \frac{\left(-1\right)^{n+m}}{n! m!} F_{p,D}^{n,m}\left(t\right)  G_{m-n,2p+n+m}^{\left(D\right)}.
\end{align}

According to \cite{Kaviraj_19} every $D-2$ dimensional conformal block can be written in terms of finitely many $D$ dimensional conformal blocks. For the following normalization

\begin{align}
    G_{J,\Delta}^{\left(D\right)}\left(z,\bar{z}\right)\underset{z,\bar{z}\xrightarrow[]{}0}{\sim} \frac{J!}{\left(-2\right)^{J}\left(\frac{D}{2}-1\right)_{J}} \left(z \bar{z}\right)^{\frac{\Delta}{2}} C^{\left(\frac{D}{2}-1\right)}_{J} \left(\frac{z+\bar{z}}{2\sqrt{z \bar{z}}}\right),
\end{align}

with $C^{\left(\frac{D}{2}-1\right)}_{J} \left(x\right)$ being the Gegenbauer polynomials, the conformal blocks can be written as

\begin{align}
    &G_{J,\Delta}^{\left(D-2\right)} = G_{J,\Delta}^{\left(D\right)}+c_{0,2}^{J,\Delta} G_{J,\Delta+2}^{\left(D\right)}+c_{-2,0}^{J,\Delta} G_{J-2,\Delta}^{\left(D\right)}+c_{-2,2}^{J,\Delta} G_{J-2,\Delta+2}^{\left(D\right)}, \\ 
    &c_{0,2}^{J,\Delta}=\frac{\Delta\left(1-\Delta\right)\left(\Delta+J\right)^{2}}{4\left(D-2\Delta-4\right)\left(D-2\Delta-2\right)\left(\Delta+J-1\right)\left(\Delta+J+1\right)}, \\
    &c_{-2,0}^{J,\Delta} = \frac{J\left(1-J\right)}{\left(D+2J-6\right)\left(D+2J-4\right)}, \\
    &c_{-2,2}^{J,\Delta} = \frac{\Delta\left(1-\Delta\right)J\left(1-J\right)\left(D-\Delta+J-4\right)^{2}\left(D-\Delta+J-5\right)^{-1}}{4\left(D-2\Delta-4\right)\left(D-2\Delta-2\right)\left(D-\Delta+J-3\right)\left(D+2J-6\right)\left(D+2J-4\right)}.
\end{align}

    

By reorganizing the terms in $\left(\ref{eq:34}\right)$ we will be left with the following expression

\begin{align} \label{eq:37}
   & \frac{1}{n!m!}F_{p,D}^{n,m}\left(t\right) = \frac{1}{n!m!}F_{p,D-2}^{n,m}\left(t\right)+ \frac{c_{0,2}^{m-n,2a+n+m-2}}{\left(n-1\right)!\left(m-1\right)!}F_{p,D-2}^{n-1,m-1}\left(t\right)\\
    \nonumber &+ \frac{c_{-2,0}^{m-n+2,2a+n+m}}{\left(n-1\right)!\left(m+1\right)!}F_{p,D-2}^{n-1,m+1}\left(t\right)+\frac{c_{-2,2}^{m-n+2,2a+n+m-2}}{\left(n-2\right)!m!}F_{p,D-2}^{n-2,m}\left(t\right), ~~ n\geq2.
\end{align}

For $n=0$ and $n=1$ we remove the terms with negative parameters. Suppose that $F_{p,D-2}^{n,m}\left(-t\right)=\left(-1\right)^{n+m}F_{p,D-2}^{n,m}\left(t\right)$, then the same is true for $F_{p,D}^{n,m}\left(t\right)$. Using the parity of $F_{p,2}^{n,m}\left(t\right)$ proven above we have parity for all even $D$. The results of \cite{Hogervorst_2016} present a much more complicated relation between the conformal blocks of dimension $D$ and $D+1$

\begin{align}
    G_{J,\Delta}^{\left(D+1\right)} = \underset{\underset{j-J~Mod2= 0}{j=0}}{\overset{J}{\sum}} \underset{n=0}{\overset{\infty}{\sum}} \mathcal{A}_{j,n}\left(J,\Delta\right) G_{j,\Delta+2n}^{\left(D\right)}.
\end{align}

The $D+1$ blocks breaks into an infinite number of $D$ blocks. This leads to a complicated relation between the $D$ and $D+1$ dimensional polynomial $F$

\begin{align}
    \frac{1}{n!m!}F_{p,D}^{n,m}\left(t\right) = \underset{\underset{\underset{k+l~Mod2=m+n~Mod2}{k-l~Mod2=0}}{k, l}}{\overset{\overset{k+l\leq m+n}{k-l\geq m-n}}{\sum}} \mathcal{A}_{m-n,\frac{m+n-k-l}{2}}\left(k-l,2a+k+l\right) \frac{1}{l!k!}F_{p,D+1}^{l,k}\left(t\right).
\end{align}

This relation defines the polynomials of dimension $D$ as complicated sums of those of dimension $D+1$. Still because $k+l-m-n \equiv 0 \mod{2}$ the $F_{p,D+1}^{n,m}\left(t\right)$ will induce parity in the $F_{p,D}^{n,m}\left(t\right)$. Although these relation allow us to write $F$ without using inversion formulas, \cite{Caron_Huot_2017,Hogervorst_2017}, for the odd dimensions expressions don't have any known closed form.

To obtain $\left(\ref{eq:16}\right)$ all that is needed is to use the parity of $F_{p,D}^{n,m}\left(t\right)$ and follow the same steps in $\left(\ref{eq:23}\right)$.

\subsection{Bound on the asymptotic}

By using a computer to check tens of thousands of coefficients and using the asymptotic behavior of $f_{p}^{n}\left(t\right)$ there is a reasonable amount of evidence for the existence of the green points in \ref{fig:03}. However, by using the correct tools we can prove their existence.

\newpage

\begin{figure}[h]
    \centering
    \includegraphics[scale=0.5]{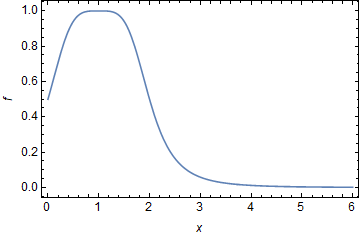}
    \caption{Let's start with a function $f\left(x\right)$ that behaves as $\frac{1}{x^{4}}$ at infinity.}
    \label{fig:05}
\end{figure}

\begin{figure}[h]

\begin{minipage}[b]{0.45\linewidth}

    \centering
    \includegraphics[scale=0.5]{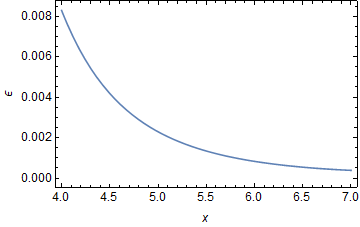}
    \caption{The error $\epsilon\left(x\right)=x^{4}f\left(x\right)-1$ between the function and its asymptotic form appears to decrease for $x\geq4$.}
    \label{fig:06}

\end{minipage} 
\hfill
\begin{minipage}[b]{0.45\linewidth}

    \centering
    \includegraphics[scale=0.5]{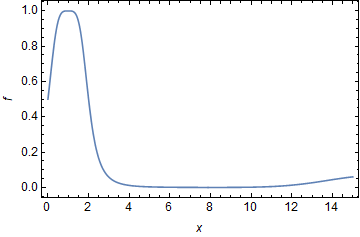}
    \caption{For $8<x<x_{0}$ the function grows, so even though the error $\epsilon\left(x\right)$ appears to decrease in the last plot we cannot guarantee it will continue to do so.}
    \label{fig:7}

\end{minipage} 

\end{figure}

Our problems stem from the asymptotic series, unlike convergent ones they only have to make sense as a limit to infinity. The first problem with these series is portrait by the function $\frac{1}{1+exp\left(\beta\left(t-\mu\right)\right)}$, it behave as $\mathcal{O}\left(e^{-\beta t}\right)$ close to infinite, but it can be delayed for an arbitrary large $t$ by changing $\mu$. The second is portrait by the three plots above, we do not know whether the error will continue to decrease without more information.

Our first tool will arise from this simple contradiction, suppose that $\underset{n\xrightarrow[]{}\infty}{\lim} \frac{g\left(n\right)}{Ag\left(n\right)}=1$, $g\left(k\right)>Ag\left(k\right)$ and for $n\geq k$, $s\left(n\right)\geq As\left(n\right)$, with $s\left(n\right)=\frac{g\left(n+1\right)}{g\left(n\right)}$ and $As\left(n\right)=\frac{Ag\left(n+1\right)}{Ag\left(n\right)}$

\begin{align}
    \frac{g\left(n\right)}{Ag\left(n\right)}=\underset{i=0}{\overset{n-k-1}{\prod}}\left(\frac{s\left(i+k\right)}{As\left(i+k\right)}\right)\frac{g\left(k\right)}{Ag\left(k\right)}\geq\frac{g\left(k\right)}{Ag\left(k\right)}>1.
\end{align}

The limit cannot be $1$. Consider the propositions $\underset{n\xrightarrow[]{}\infty}{\lim} \frac{g\left(n\right)}{Ag\left(n\right)}=1$ and for $n\geq k$, $s\left(n\right)\geq As\left(n\right)$ to be true, then $\epsilon\left(n\right)=1-\frac{g\left(n\right)}{Ag\left(n\right)}>0$ for $n\geq k$, consider $n_{1}>n_{2}$

\begin{align}
    \frac{g\left(n_{1}\right)}{Ag\left(n_{1}\right)}=\underset{i=0}{\overset{n_{1}-n_{2}-1}{\prod}}&\left(\frac{s\left(i+n_{2}\right)}{As\left(i+n_{2}\right)}\right)\frac{g\left(n_{2}\right)}{Ag\left(n_{2}\right)}\geq\frac{g\left(n_{2}\right)}{Ag\left(n_{2}\right)} \\ 
    1-\epsilon\left(n_{1}\right)&\geq1-\epsilon\left(n_{2}\right) \Longrightarrow \epsilon\left(n_{1}\right)\leq \epsilon\left(n_{2}\right).
\end{align}

With this we can prove the error between the function and its asymptotic expression strictly decreases. Note that all of these and the following arguments are still true if we change the inequality signs in the contradiction and define $\epsilon\left(n\right)$ as $\frac{g\left(n\right)}{Ag\left(n\right)}-1$.

Because
 of the existence of the $\left(-1\right)^{n}$ in $\left(\ref{eq:55}\right)$ it is better to treat even and odd $n$ separately. Without a simple expression for the $f_{p}^{n}\left(t\right)$, which coincidentally exist for integer and half-integer $t$, we cannot directly use the arguments above. For general $t$ we rewrite the recursion relation $\left(\ref{eq:22}\right)$ in terms of $s_{p}^{n}\left(t\right)=\frac{f_{p}^{n+1}\left(t\right)}{\left(n+1\right)f_{p}^{n}\left(t\right)}$, and similar for $As_{p}^{n}\left(t\right)$

\begin{align}
    & s_{p}^{n}\left(t\right)=\frac{t}{n+1}+\dfrac{\left(n+p-1\right)^{2} \left(n+2p-2\right)}{4\left(n+1\right)\left(2n+2p-1\right)\left(2n+2p-3\right)s_{p}^{n-1}\left(t\right)}, \\ 
    \label{eq:58} \Longrightarrow & s_{p}^{n+1}\left(t\right)=\frac{t+\frac{s_{p}^{n-1}\left(t\right)\left(n+1\right)\left(p+n\right)^{2}\left(n+2p-1\right)\left(2n+2p-3\right)}{\left(2n+2p+1\right)\left(\left(n+p-1\right)^{2}\left(n+2p-2\right)+4t~\!\!s_{p}^{n-1}\left(t\right)\left(2n+2p-3\right)\left(2n+2p-1\right)\right)}}{n+2}.
\end{align}

This expression has two important properties. First, it grows monotonically, if $s_{p}^{n-1}\left(t\right)>r_{p}^{n-1}\left(t\right)$ are two initial conditions for $\left(\ref{eq:58}\right)$ then $s_{p}^{n+1}\left(t\right)>r_{p}^{n+1}\left(t\right)$. Second, it is attractive, there is a real positive solution  $s^{*}\left(n,p,t\right)$ to the equation $s_{p}^{n-1}\left(t\right)=s_{p}^{n+1}\left(t\right)$ and if $s_{a}^{n-1}\left(t\right)> s^{*}\left(n,p,t\right)$, or $s_{a}^{n-1}\left(t\right)< s^{*}\left(n,p,t\right)$, then $s_{a}^{n-1}\left(t\right)>s_{a}^{n+1}\left(t\right)$, or $s_{a}^{n-1}\left(t\right)<s_{a}^{n+1}\left(t\right)$. These means that the flow in the phase space contracts around the curve $s^{*}\left(n,p,t\right)$.

To prove the inequality $s_{p}^{n}\left(t\right)\geq As_{p}^{n}\left(t\right)$ for $n\geq k$ we start by finding a curve $b\left(n\right)$ where the flow always points to the region above $b$. If we find a point $k_{1}$ such that $s_{p}^{k_{1}}\left(t\right)>b\left(k_{1}\right)$ we would know that it continues to be true for $n\geq k_{1}$, if $b\left(n\right)\geq As_{p}^{n}\left(t\right)$ for $n\geq k_{2}$ we would have indirectly proven that $s_{p}^{n}\left(t\right)\geq As_{p}^{n}\left(t\right)$ for $n\geq k=\max\left[k_{1},k_{2}\right]$. However even if we found the biggest $k$ between all of the $f_{p}^{n}\left(t\right)$ it would not enough for the $\mathcal{E}_{\left[p\right]}^{n,m}$.

Imagine that $f\left(n\right)$ approaches $10$ while $g\left(n\right)$ approaches $-9$, even when both have an error within $1\%$ the bound for the sum $f+g$ still has only $19\%$ accuracy, $0.81\leq f\left(n\right)+g\left(n\right)\leq 1.19$. The same problems occur in every $\mathcal{E}_{\left[p\right]}^{n,m}$ when dealing with non-trivial solutions.

Although we cannot directly bound the errors for the $\mathcal{E}_{\left[p\right]}^{n,m}$ using the bounds form the $f_{p}^{n}\left(t\right)$ we still can use them to construct a lower bound $L\mathcal{E}_{\left[p\right]}^{n,m}$ and prove its positivity. The exact map between $\mathcal{E}_{\left[p\right]}^{n,m}$ and $L\mathcal{E}_{\left[p\right]}^{n,m}$ is quite complicated and not particularly illuminating so we will refer back to the program\footnote{The program can be found at \href{https://github.com/rgnato/Generalized_polynomials}{GitHub}.}. What is important is that we can find a value $n^{*}$ and $J^{*}\left(n\right)$ such that for $n\geq n^{*}$ or $J\geq J^{*}\left(n\right)$ we can guarantee the positivity of every $\mathcal{E}_{\left[p\right]}^{n,m}$. To prove unitarity we just need to test the cases $n<n^{*}$ and $J<J^{*}\left(n\right)$.

For the special case in figure $\ref{fig:03}$ we can use analytic properties of the continuous Hahn polynomials to find a similar lower bound and skip need to deal with the asymptotic expressions. Again the proof of this statement is long so we won't show here.


\begin{thebibliography}{10}

\bibitem{FERRARA1973161}
S. Ferrara, A. Grillo, and R. Gatto, ``Tensor representations of conformal
  algebra and conformally covariant operator product expansion,'' Annals of
  Physics {\bf 76}, 161  (1973).

\bibitem{Polyakov:1974gs}
A. Polyakov, ``{Nonhamiltonian approach to conformal quantum field theory},''
  Zh. Eksp. Teor. Fiz. {\bf 66}, 23 (1974).

\bibitem{Rattazzi_2008}
R. Rattazzi {\it et~al.}, ``Bounding scalar operator dimensions in 4DCFT,''
  Journal of High Energy Physics {\bf 2008}, 031 (2008).

\bibitem{El_Showk_2012}
S. El-Showk {\it et~al.}, ``Solving the 3D Ising model with the conformal
  bootstrap,'' Physical Review D 86 (2012).

\bibitem{Kos_2014}
F. Kos, D. Poland, and D. Simmons-Duffin, ``Bootstrapping mixed correlators in
  the 3D Ising model,'' Journal of High Energy Physics 2014 (2014).

\bibitem{Kos_2015}
F. Kos {\it et~al.}, ``Bootstrapping the O(N) archipelago,'' Journal of High
  Energy Physics 2015 (2015).

\bibitem{Poland:2018epd}
D. Poland, S. Rychkov, and A. Vichi, ``{The Conformal Bootstrap: Theory,
  Numerical Techniques, and Applications},'' Rev. Mod. Phys. {\bf 91}, 015002
  (2019).

\bibitem{Rychkov:2016iqz}
S. Rychkov, {\it {EPFL Lectures on Conformal Field Theory in D\ensuremath{>}= 3
  Dimensions}}, {\it SpringerBriefs in Physics} (2016).

\bibitem{simmonsduffin2016tasi}
D. Simmons-Duffin, ``TASI Lectures on the Conformal Bootstrap,'', 2016.

\bibitem{2020}
A.~L. Guerrieri, A. Homrich, and P. Vieira, ``Dual S-matrix bootstrap. Part I.
  2D theory,'' Journal of High Energy Physics 2020 (2020).

\bibitem{Hogervorst_2017}
M. Hogervorst and B.~C. van Rees, ``Crossing symmetry in alpha space,'' Journal
  of High Energy Physics 2017 (2017).

\bibitem{Paulos_2017}
M.~F. Paulos {\it et~al.}, ``The S-matrix bootstrap. Part I: QFT in AdS,''
  Journal of High Energy Physics 2017 (2017).

\bibitem{Hogervorst_2016}
M. Hogervorst, ``Dimensional reduction for conformal blocks,'' Journal of High
  Energy Physics 2016 (2016).

\bibitem{Kaviraj_19}
A. Kaviraj, S. Rychkov, and E. Trevisani, ``Random Field Ising Model and
  Parisi-Sourlas supersymmetry. Part I. Supersymmetric CFT,'' Journal of High
  Energy Physics {\bf 2020}, 90 (2020).

\bibitem{book}
R. Koekoek, P. Lesky, and R. Swarttouw, {\it Hypergeometric Orthogonal
  Polynomials and Their q-Analogues} (2010).

\bibitem{Caron_Huot_2017}
S. Caron-Huot, ``Analyticity in spin in conformal theories,'' Journal of High
  Energy Physics 2017 (2017).

\end{thebibliography}

\end{document}